\documentclass[jpd,superscriptaddress,twocolumn,showpacs]{revtex4-1}
\usepackage{amsmath,graphicx,bm,mathtools}
\usepackage[colorlinks,linkcolor=blue,anchorcolor=green,citecolor=red]{hyperref}

\newcommand{\heff}{\bm H_\text{eff}}

\begin{document}
\title{Anomalous Feedback and Negative Domain Wall Resistance}

\author{Ran Cheng}
\affiliation{Department of Physics, Carnegie Mellon University, Pittsburgh, PA 15213, USA}

\author{Jian-Gang Zhu}
\affiliation{Department of Electrical and Computer Engineering, Carnegie Mellon University, Pittsburgh, PA 15213, USA}

\author{Di Xiao}
\affiliation{Department of Physics, Carnegie Mellon University, Pittsburgh, PA 15213, USA}

\begin{abstract}

Magnetic induction can be regarded as a negative feedback effect, where the motive-force opposes the change of magnetic flux that generates the motive-force. In artificial electromagnetics emerging from spintronics, however, this is not necessarily the case. By studying the current-induced domain wall dynamics in a cylindrical nanowire, we show that the spin motive-force exerting on electrons can either oppose or support the applied current that drives the domain wall. The switching into the anomalous feedback regime occurs when the strength of the dissipative torque $\beta$ is about twice the value of the Gilbert damping constant $\alpha$. The anomalous feedback manifests as a negative domain wall resistance, which has an analogy with the water turbine.

\end{abstract}

\maketitle

\section{Introduction}

Magnetization dynamics and electron transport are coupled together in a reciprocal manner. Their interplay introduces a variety of feedback phenomena~\cite{ref:volovik,ref:JDZang,ref:Schulz,ref:SVfeedback1,ref:JXfeedback,ref:Foros,ref:Zhang,ref:Hydro,ref:heatpump,ref:SVfeedback2,ref:feedback,ref:AFoscillator}. For example, when a background magnetization varies slowly over space and time, conduction electron spins will follow the magnetization orientation. By doing so, the electron wave function acquires a geometric phase changing with time, which behaves as a time-varying magnetic flux and produces a spin motive-force (SMF) according to Faraday's effect~\cite{ref:SMF,ref:Shengyuan}. As a feedback, electrons driven by the SMF react on the magnetization via the spin-transfer torque (STT)~\cite{ref:STTorig,ref:Bazaliy,ref:STT}. This reaction leads to a modified magnetic damping, which hinders the magnetization dynamics that generates the SMF~\cite{ref:Zhang}. In parallel, when a magnetic texture is driven into motion by a current, it in turn exerts SMFs on the electrons, resulting in a modified electrical resistivity that inhibits the growth of the driving current~\cite{ref:JDZang,ref:Schulz}. 

Similar feedback mechanisms also apply to magnetic heterostructures~\cite{ref:feedback}. For example, spin current pumped from a precessing ferromagnet into an adjacent normal metal experiences a backflow, which, in turn acts on the ferromagnet through STT~\cite{ref:Bauer}. Because of the backflow-induced STT, the effective spin-mixing conductance on the interface is renormalized~\cite{ref:JX}. If the pumped spin current is absorbed by a second ferromagnet instead of flowing back, it will mediate a dynamical interlayer coupling between the two ferromagnets~\cite{ref:SVfeedback1,ref:SVfeedback2}. Recently, it has also been shown that in the presence of the spin Hall effect, spin pumping and spin-backflow are connected through a feedback loop due to the combined effect of the spin Hall and its reverse process~\cite{ref:feedback,ref:AFoscillator}. This novel feedback mechanism, despite quadratic in the spin Hall angle, gives rise to a crucial nonlinear damping effect that qualitatively changes the dynamical behavior of the magnetization.

In electromagnetics, a negative feedback is ensured by the Lenz law~\cite{ref:Lenz}, which requires that the emf generated by Faraday's effect must oppose the change of magnetic flux that causes the emf. For instance, an electric motor works simultaneously as a dynamotor so that the induced emf counteracts the applied emf. As a result, the electric current flowing through its coil is attenuated and the resistance from $I-V$ measurement is larger than the resistance of the coil. In the context of spintronics, the current-induced magnetization dynamics plays the role of an electric motor, which in turn drives the current in a similar fashion as a dynamotor. Regarding the Lenz law, one may expect an increased resistivity.

In this paper, however, we show that this naive expectation is not always correct. The feedback acting on the driving current can also give rise to a reduced resistivity. As an example, we study the current-driven domain wall (DW) dynamics in a nanowire with cylindrical symmetry~\cite{ref:cylinder}, and demonstrate that when the DW is set into motion by an applied current, its reaction in the form of SMF can either propel or repel the electron motion, creating either a negative or a positive DW resistance. The sign of the DW resistance reflects the style of the feedback, which depends only on two phenomenological parameters---the Gilbert damping constant $\alpha$ and the strength of the dissipative torque $\beta$. To interpret such an anomalous feedback phenomenon, we make an analogy to the working mechanism of a water turbine. It is observed that if a DW propels electrons along with its motion, just like a rotating turbine wheel carriers water, a negative DW resistance is produced.

The paper is organized as follows. In Sec.~\ref{feedback}, we establish the general formalism. In Sec.~\ref{spintexture}, we apply the formalism to a slowly-varying spin texture and derive the feedback-induced change of dissipations. In Section~\ref{DW}, we explore the current-driven DW dynamics in a cylindrically symmetric nanowire, and derive the DW resistance in terms of $\alpha$ and $\beta$. In Section~\ref{disc}, we provide an intuitive interpretation of the anomalous feedback.

\section{Dynamic feedbacks}\label{feedback}

\begin{figure}[t]
	\centering
	\includegraphics[width=0.95\linewidth]{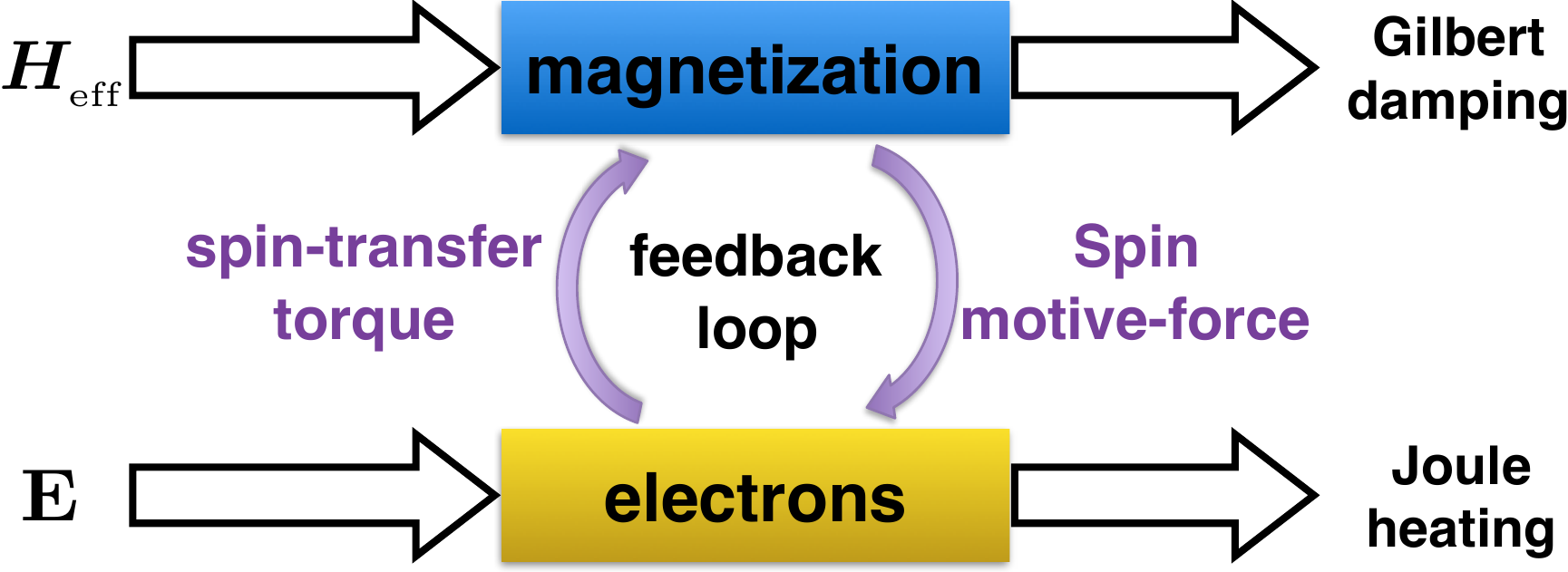}
	\caption{(Color online) The interplay between magnetization and conduction electrons generates a dynamic feedback loop that connects the magnetic and electronic dissipations.}
	\label{fig:loop}
\end{figure}

As illustrated in Fig.~\ref{fig:loop}, the interplay between local magnetization and conduction electrons is resolved in a dynamic feedback loop connecting energy dissipation channels of each individual process. Under the adiabatic assumption~\cite{ref:DX}, we regard the magnetic order parameter $\bm{m}(\bm{r},t)$ as a slowly-varying vector in space and time so that conduction electron spins are able to adjust to the magnetization direction. Given the magnetic free energy $\mathcal{U}[\bm{m}(\bm{r},t)]$, we define the effective magnetic field as $\heff=-\delta\mathcal{U}/\delta\bm{m}$. In the diffusive region, nonlocal processes are suppressed, and the coupled dynamics of the system is described by
\begin{subequations}
	\label{eq:couple}
	\begin{align}
	(1-&\hat{\alpha}\bm{m}\times)\dot{\bm{m}}=\gamma\heff\times\bm{m}+\bm{\tau}(\bm{j}), \label{eq:mdot}\\
	\bm{j}&=\hat{G}(\bm{m})\bm{E}+\bm{\varepsilon}(\dot{\bm{m}}), \label{eq:j}
	\end{align}
\end{subequations}
where $\gamma$ is the gyromagnetic ratio, $\hat{\alpha}$ is the magnetic damping tensor, $\hat{G}(\bm{m})$ is the conductivity tensor. The STT $\bm{\tau}$ and the motive force $\bm{\varepsilon}$ are \textit{local} functions of $\bm{j}$ and $\dot{\bm{m}}$, respectively; they mix the dynamics of $\bm{m}$ with that of electrons. Note that $\bm{\tau}$ and $\bm{\varepsilon}$ may also depend on the spatial gradient of the magnetization $\nabla\bm{m}$. With proper initial conditions, the evolution of $\bm{m}$ and $\bm{j}$ can be solved by iterations of Eq.~\eqref{eq:couple} on discretized spacetime grid. At any particular point $(\bm{r},t)$, one is allowed to eliminate $\bm{j}$ (or $\dot{\bm{m}}$) by substituting Eq.~\eqref{eq:j} into Eq.~\eqref{eq:mdot} [or Eq.~\eqref{eq:mdot} into Eq.~\eqref{eq:j}] if both $\bm{\tau}$ and $\bm{\varepsilon}$ are local functions of the space and time coordinates.

Such an elimination operation fulfills the feedback loop illustrated in Fig.~\ref{fig:loop}. For example, if $\bm{E}=0$, the current $\bm{j}$ is only induced by the motion of $\bm{m}$ through $\bm{\varepsilon}$, which is simultaneously reacting on $\bm{m}$ by virtue of $\bm{\tau}$. In this regard, we can eliminate $\bm{j}$ by inserting Eq.~\eqref{eq:j} into Eq.~\eqref{eq:mdot}, which modifies the magnetic damping tensor $\hat{\alpha}$. In a parallel sense, if the magnetization dynamics is solely driven by $\bm{j}$ (no magnetic field), it also generates a feedback on $\bm{j}$ and renormalizes the conductivity tensor $\hat{G}$. The latter corresponds to the elimination of $\dot{\bm{m}}$ by inserting Eq.~\eqref{eq:mdot} into Eq.~\eqref{eq:j}.

The dynamic feedback effects can be further elucidated by energy dissipations. Swapping the roles of the thermodynamic forces $\heff$ and $\bm{E}$ with the corresponding currents $\dot{\bm{m}}$ and $\bm{j}$~\cite{note0}, we can rewrite Eq.~\eqref{eq:couple} as
\begin{align}
\begin{bmatrix}
\heff \\ \bm{E}
\end{bmatrix}
=
\begin{bmatrix}
L_{11} & L_{12} \\
L_{21} & L_{22} 
\end{bmatrix}
\begin{bmatrix}
\dot{\bm{m}} \\ \bm{j}
\end{bmatrix}.
\label{eq:Lmatrix}
\end{align}
Here, $L_{11}$ is pertaining to the Gilbert damping, $L_{12}$ the current-induced torque, $L_{21}$ the motive force, and $L_{22}$ the electrical resistivity. The Onsager's reciprocity relation implies that $L^{\mathrm{T}}_{12}(\bm{m},\heff)=L_{21}(-\bm{m},-\heff)$~\cite{ref:Landau}. If magnetization and current decouple, \textit{i.e.}, $L_{12}=0$, the magnetic free energy dissipates only through the Gilbert damping $\dot{\mathcal{U}}_m=-\heff\cdot\dot{\bm{m}}=-L_{11}\dot{\bm{m}}^2$, while the electron free energy dissipates only through the Joule heating $\dot{\mathcal{U}}_e=-\bm{E}\cdot\bm{j}=-L_{22}\bm{j}^2$. However, when the STT ($L_{12}$) and the motive force ($L_{21}$) are introduced, a feedback loop will connect the two channels of energy dissipation as shown in Fig.~\ref{fig:loop}. For example, the magnetic dissipation is implemented by not only the Gilbert damping, but also the Joule heating, since a magnetic precession inevitably drives the electron motion that carries away the magnetic energy and subsequently dissipates into heat. This manifests as a renormalization of the magnetic damping tensor $\hat{\alpha}$ (thus $L_{11}$). In a similar fashion, electron current can excite magnetic precession, which takes away the electron kinetic energy and damped into heat through the Gilbert damping. As a result, the resistivity tensor $L_{22}$ is effectively modified. The rates of free energy loss are thus
\begin{subequations}
	\label{eq:L1122}
	\begin{align}
	\dot{\mathcal{U}}_m&=-\dot{\bm{m}}\left[L_{11}-L_{12}L_{22}^{-1}L_{21}\right]\dot{\bm{m}}\equiv-\mathcal{L}_{11}\dot{\bm{m}}^2, \\
	\dot{\mathcal{U}}_e&=-\bm{j}\left[L_{22}-L_{21}L_{11}^{-1}L_{12}\right]\bm{j}\equiv-\mathcal{L}_{22}\bm{j}^2,
	\end{align}
\end{subequations}
where $\mathcal{L}_{11}$ and $\mathcal{L}_{22}$ are the response coefficients modified by the dynamic feedback.

In general, if a system is driven by a set of $N$ thermodynamic forces [or currents in the ``swapped'' convention, see Eq.~\eqref{eq:Lmatrix}] $X_1$, $X_2$, $\cdots X_N$, there are $N$ currents (or forces) $J_1$, $J_2$, $\cdots J_N$ satisfying $J_a=L_{ab}X_b$, where the repeated index is summed. By a straightforward derivation elaborated in the Appendix, the renormalized energy dissipation rate through a particular channel $k$ is
\begin{align}
\dot{U}_k=-\frac{X_k^2}{[L^{-1}]_{kk}}, \label{eq:Linv}
\end{align}
where $L^{-1}$ denotes the inverse of the response matrix. For $N=2$, Eq.~\eqref{eq:Linv} reduces to Eq.~\eqref{eq:L1122}. We mention that Eq.~\eqref{eq:Linv} is quite general, where the thermodynamic forces (or currents) can be magnetic, electric, thermalic, \textit{etc}. However, to simplify the following discussions, we do not include any thermoelectric effect, although they may become important in many circumstances~\cite{ref:heatpump}.

\section{Spin texture}\label{spintexture}

\subsection{Damping}

As mentioned earlier, a spacetime dependent magnetization $\bm{m}(\bm{r},t)$ drives local spin currents via the SMF. The SMF that exerts on spin-up electrons is opposite to its counterpart that exerts on spin-down electrons: $\bm{\varepsilon}^{\uparrow}=-\bm{\varepsilon}^{\downarrow}$, where the spin direction is determined with respect to the local and instantaneous $\bm{m}(\bm{r},t)$. Since the spin current is polarized along $\bm{m}$, we only keep its flow direction in the subscript, so the $i$-component of the spin current density is
\begin{align}
j_i^s&=\frac{\mu_B}{e}(G_{ik}^{\uparrow}\varepsilon_k^{\uparrow}-G_{ik}^{\downarrow}\varepsilon_k^{\downarrow}) \notag\\
&=\frac{\mu_B\hbar G^c_{ik}}{2e^2}[(\partial_t\bm{m}\times\partial_k\bm{m})\cdot\bm{m}+\beta\partial_t\bm{m}\cdot\partial_k\bm{m}], \label{eq:js}
\end{align}
where $G_{ik}^c=G^{\uparrow}_{ik}+G^{\downarrow}_{ik}$ is the $ik$-component of the conductivity tensor, $\mu_B$ is the Bohr magneton, and the Land{\'e} $g$-factor of electrons is taken to be 2. The term proportional to $\beta$ is the dissipative SMF~\cite{ref:Duinebeta,ref:Yaroslavbeta}, which is the reciprocal effect of the dissipative STT; $\beta$ is a phenomenological constant that characterizes the relative strengths of the dissipative contribution.

As a feedback effect, the locally pumped spin current acts on the magnetization through the STT. Define the electron velocity field as  $\bm{u}=\bm{j}^s/M_s$, where $M_s$ is the saturation magnetization. Then the STT consists of two orthogonal terms~\cite{ref:STT}
\begin{align}
\bm{\tau}=(u_i\partial_i)\bm{m}-\beta\bm{m}\times(u_i\partial_i)\bm{m}. \label{eq:taus}
\end{align}
Inserting Eq.~\eqref{eq:js} into Eq.~\eqref{eq:taus} yields a damping term that renormalizes the original Gilbert damping. The Landau-Lifshitz-Gilbert (LLG) equation becomes
\begin{align}
\partial_t\bm{m}=\gamma\bm{H}_{\mathrm{eff}}\times\bm{m}+\bm{m}\times(\mathcal{D}\cdot\partial_t\bm{m}),
\end{align}
where $\mathcal{D}$ is the damping tensor that can be decomposed into $\hat{\mathcal{D}}=\hat{\mathcal{D}}_0+\hat{\mathcal{D}}'$, where $\hat{\mathcal{D}}_0=\alpha_0\mathrm{\hat{\bm{I}}}$ is the original Gilbert damping, and the feedback correction is
\begin{align}
\hat{\mathcal{D}}'=\eta[\hat{\mathcal{S}}+\hat{\mathcal{A}}] \label{eq:Dab}
\end{align}
with $\eta=\mu_B\hbar/(2e^2M_s)$. In Eq.~\eqref{eq:Dab}, the element of the symmetric part is
\begin{align}
\mathcal{S}_{ab}=G_{ik}^c&\left[(\bm{m}\times\partial_i\bm{m})_a(\bm{m}\times\partial_k\bm{m})_b \right. \notag\\
&\qquad\left.-\beta^2(\partial_i\bm{m})_a(\partial_k\bm{m})_b\right],
\end{align}
and that of the antisymmetric part is
\begin{align}
\mathcal{A}_{ab}=\beta G_{ik}^c\left[(\partial_i\bm{m})_a(\bm{m}\times\partial_k\bm{m})_b-(a\rightleftharpoons b)\right], \label{eq:Aab}
\end{align}
where summations over repeated indices are assumed. In matrix form, the feedback correction can be written as $\hat{\mathcal{D}}'=\eta\ \mathcal{T}_{_{\mathrm{STT}}}\otimes\mathcal{T}_{_{\mathrm{SMF}}}=\eta G_{ik}^c[(\bm{m}\times\partial_i\bm{m})+\beta\partial_i\bm{m}]\otimes[(\bm{m}\times\partial_k\bm{m})-\beta\partial_k\bm{m}]$. This suggestive form interprets the feedback loop as two combined processes: a dynamic $\bm{m}$ pumps a local spin current, which in turn acts on $\bm{m}$ itself, implementing the feedback effect. When $\beta\rightarrow0$, Eq.~\eqref{eq:Dab} reduces to Eq.~(11) in Ref.~\cite{ref:Zhang}.

Here is an important remark. Although equations~\eqref{eq:Dab}--\eqref{eq:Aab} are similar to the results derived in Ref.~\cite{ref:Foros,ref:Hydro}, the underlying physics is fundamentally distinct. In Ref.~\cite{ref:Foros,ref:Hydro}, the damping renormalization is attributed to the current-induced noise, and thermal fluctuation is the primary stimulus. Consequently, the coefficient of the damping tensor depends on temperature. By contrast, our results are valid even at zero temperature.

\subsection{Resistance}

When closing the feedback loop the other way around, \textit{i.e.}, current $\xrightarrow[]{\rm STT}$ LLG $\xrightarrow[]{\rm SMF}$ current, we will obtain the feedback modification of the resistance. To perform this calculation, we start with the LLG equation
\begin{align}
\partial_t\bm{m}=\gamma\bm{H}_{\mathrm{eff}}\times\bm{m}&+\alpha\bm{m}\times\partial_t\bm{m} \notag\\
&+(u_i\partial_i)\bm{m}-\beta\bm{m}\times(u_i\partial_i)\bm{m},
\end{align}
then combine all $\partial_t\bm{m}$ terms so that
\begin{align}
\partial_t\bm{m}&=\frac{\gamma}{1+\alpha^2}[\bm{H}_{\mathrm{eff}}\times\bm{m}+\alpha\bm{m}\times(\bm{H}_{\mathrm{eff}}\times\bm{m})] \notag\\
&\ +\frac{1+\alpha\beta}{1+\alpha^2}(u_i\partial_i)\bm{m}+\frac{\alpha-\beta}{1+\alpha^2}\bm{m}\times(u_i\partial_i)\bm{m}, \label{eq:partt}
\end{align}
where $\bm{u}=P\mu_B\bm{j}^c/(eM_s)$ with $P=(n_{\uparrow}^F-n_{\downarrow}^F)/(n_{\uparrow}^F+n_{\downarrow}^F)$ the polarization of carrier density at the Fermi level. The charge current density is now driven by both the SMF and an external electric field $\bm{E}$,
\begin{align}
j^c_i=& j_i^{c(E)}+j_i^{c(\mathrm{smf})}=\ G_{ik}^cE_k \notag\\
+&\ G_{ik}^s\frac{\hbar}{2e}[(\partial_t\bm{m}\times\partial_k\bm{m})\cdot\bm{m}+\beta(\partial_t\bm{m}\cdot\partial_k\bm{m})], \label{eq:jee}
\end{align}
where $G_{ik}^s=G^{\uparrow}_{ik}-G^{\downarrow}_{ik}$ is the $ik$-component of the spin conductivity. It should not be confused that for the SMF-induced electron flow, the spin current depends on the charge conductivity [see Eq.~\eqref{eq:js}], whereas the charge current depends on the spin conductivity~\cite{ref:Zhang}.

When substituting the LLG equation into the SMF to eliminate $\partial_t\bm{m}$, terms involving $\bm{H}_{\mathrm{eff}}$ result in nonlinear dependence between $\bm{j}_c$ and $\bm{E}$, which in principle should be solved \textit{numerically}. Nevertheless, those terms can be discarded in many special cases. For instance, if the magnetic free energy is invariant under a particular motion of $\bm{m}$, we have $\bm{H}_{\mathrm{eff}}\parallel\bm{m}$ at all times, thus those terms vanish identically. In such circumstances, $\bm{E}$ is linear in $\bm{j}_c$, and the feedback can be expressed \textit{analytically} as a renormalization of the resistivity tensor. We will restrict the following discussion to this category.

To proceed, we insert Eq.~\eqref{eq:partt} into Eq.~\eqref{eq:jee} and make the approximation that $\bm{H}_{\mathrm{eff}}\parallel\bm{m}$. After some manipulations, we obtain
\begin{align}
j^c_i+G_{ik}^s\mathcal{R}_{k\ell}j^c_\ell=G_{ik}^cE_k, \label{eq:jefeed}
\end{align}
where the element of the feedback matrix $\hat{\mathcal{R}}$ is
\begin{align}
\mathcal{R}_{k\ell}&=\frac{P\mu_B\hbar}{2e^2M_s}\left[\frac{\alpha(1-\beta^2)-2\beta}{1+\alpha^2}\partial_k\bm{m}\cdot\partial_\ell\bm{m}  \right. \notag\\
&\qquad\qquad\quad\left.+\frac{1+2\alpha\beta-\beta^2}{1+\alpha^2}(\partial_k\bm{m}\times\partial_\ell\bm{m})\cdot\bm{m} \right] \notag\\
&\equiv \frac{P\mu_B\hbar}{2e^2M_s}[f(\alpha,\beta)g_{k\ell}+h(\alpha,\beta)\Omega_{k\ell}]. \label{eq:rkl}
\end{align}
The symmetric part of $\hat{\mathcal{R}}$ is proportional to the quantum metric $g_{k\ell}=\partial_k\bm{m}\cdot\partial_\ell\bm{m}$~\cite{ref:QGT}, while the antisymmetric part is proportional to the Berry curvature $\Omega_{k\ell}=(\partial_k\bm{m}\times\partial_\ell\bm{m})\cdot\bm{m}$. To appreciate the physical meaning of $\hat{\mathcal{R}}$, we turn to the resistivity by multiplying $[\hat{G}^{c}]^{-1}$ on Eq.~\eqref{eq:jefeed}, which gives $\bm{E}=\hat{\rho}\bm{j}^c$. The resistivity tensor is
\begin{align}
 \hat{\rho}=\hat{\rho}_0(1+\hat{G}^s\hat{\mathcal{R}}),
\end{align}
where $\hat{\rho}_0=[\hat{G}^{c}]^{-1}$ is the bare resistivity tensor without feedback, and $\hat{G}^s\hat{\mathcal{R}}$ is the feedback-induced renormalization. Depending on the spatial pattern of $\bm{m}(\bm{r},t)$ and the relative ratio between $\alpha$ and $\beta$, a particular element of $\hat{\mathcal{R}}$ can be either positive or negative.

\section{Domain wall resistance}\label{DW}

Transverse DWs in thin cylindrical magnetic nanowires have two salient features that arouse recent interest~\cite{ref:cylinder}. (1) The inner structure of these DWs remain unchanged during their propagation, thus our assumption $\bm{H}_{\mathrm{eff}}\parallel\bm{m}$ is respected at all times. (2) These DWs are massless and the critical currents required to initiate their motions are zero. Because of the latter property, the DW resistance practically measurable from I-V curve solely stems from the dynamic feedback effect, whereas the conventional theory based on stationary DW configurations~\cite{ref:LevyZhang,ref:DWresistance} is incomplete.

Such a DW is a one-dimensional soliton characterized by two spherical angles $\theta$ and $\phi$ specifying the local orientation of the magnetization
\begin{subequations}
	\label{eq:profile}
	\begin{align}
		\theta(x,t)&=2\arctan e^{[x-x_c(t)]/w}, \\
		\phi(x,t)&=\phi(t),
	\end{align}
\end{subequations}
where $x_c(t)$ is the center of the DW, and $w$ is the width of the DW (supposed to be much larger than the lattice spacing). In one dimensions, the antisymmetric part of Eq.~\eqref{eq:rkl} vanishes, $\hat{\Omega}=0$; $\hat{\mathcal{R}}$ has only one component, and $G^s=PG^c$. In this case, Eq.~\eqref{eq:jefeed} reduces to $\rho\bm{j}^c=\bm{E}$, where $\rho=\left[\rho_0+P^2\eta f(\alpha,\beta)|\partial_x\bm{m}|^2\right]$ with $\eta=\mu_B\hbar/(2e^2M_s)$. The profile function given by Eq.~\eqref{eq:profile} yields $|\partial_x\bm{m}|^2=1/[w^2\cosh^2(x/w)]$. By integrating $\rho$ over $x\in(-\infty,+\infty)$, we obtain the total resistance
\begin{align}
	R=R_0+\frac{\alpha(1-\beta^2)-2\beta}{1+\alpha^2}\left[\frac{P^2\mu_B\hbar}{e^2M_s}\right]\frac1{Aw}, \label{eq:DWresistance}
\end{align}
where $A$ is the area of the cross section of the cylindrical nanowire. The second term of Eq.~\eqref{eq:DWresistance} is ascribed to the dynamic feedback effect, which scales inversely with $w$. Since $P^2\mu_B\hbar/(e^2M_sAw)>0$, the sign of this correction is only determined by $f(\alpha,\beta)=[\alpha(1-\beta^2)-2\beta]/(1+\alpha^2)$. Consider $\alpha\ll1$ and $\beta\ll1$, then $f(\alpha,\beta)\approx\alpha-2\beta$. As a result, the dynamical correction of the DW resistance is positive for $\beta<\alpha/2$, and negative for $\beta>\alpha/2$. Using typical material parameters of permalloy, the feedback-induced resistance of a 100nm wide DW with $A\sim$30nm$^2$ is in the range of $10^{-5}$ to $10^{-4}$ $\Omega$.

A negative DW resistance indicates that the feedback exerting on the electrons by the DW is positive. To be specific, when the DW is set into motion by a current, it propels the electrons in their direction of motion, thus reducing the electrical resistance. In terms of the Lenz law, this means that the SMF induction enhances the flux (geometric phase) change by making the electrons more mobile, contrasting to the normal case where the SMF opposes the flux change. It worths emphasizing that such an anomalous situation is unique to cylindrically symmetric nanowires, while nanostrips are not applicable as the approximation $\bm{H}_{\mathrm{eff}}\parallel\bm{m}$ is invalid.

\section{Discussion}\label{disc}

\begin{figure}[b]
	\centering
	\includegraphics[width=0.98\linewidth]{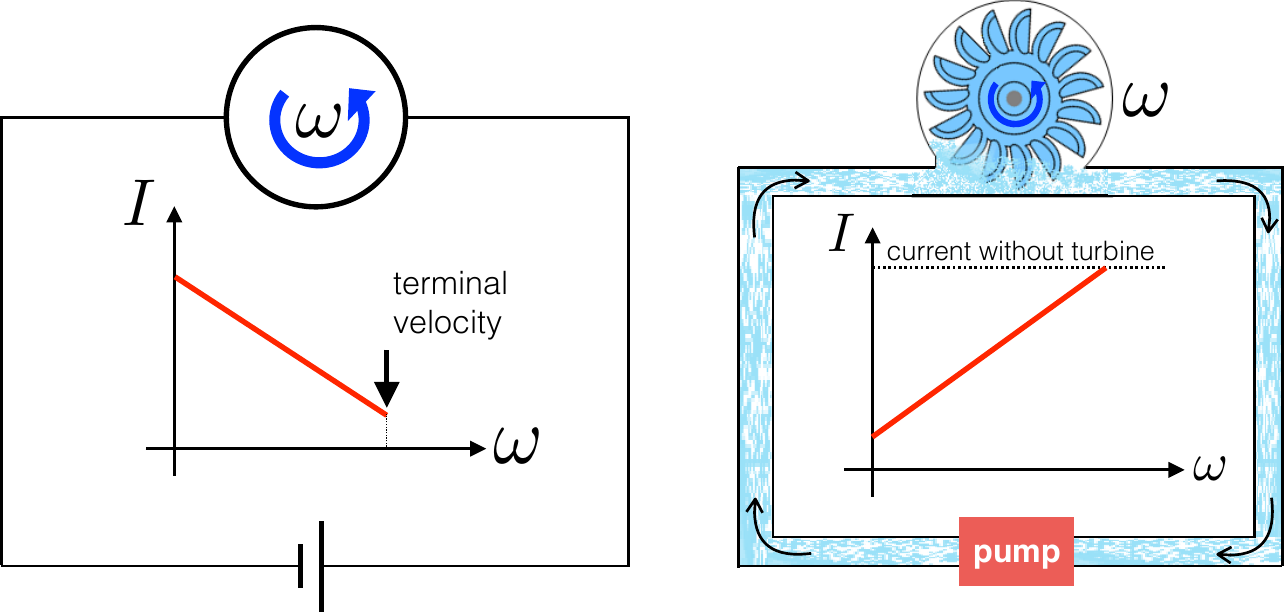}
	\caption{(Color online) Comparison between an electric motor driven by a constant voltage and a water turbine driven by a constant pump. The overall current $I$ as a function of the angular velocity $\omega$ signals the nature of the feedback effect.}
	\label{fig:machine}
\end{figure}

Different from the static DW resistance~\cite{ref:LevyZhang,ref:DWresistance} that is absorbed by $R_0$ in our theory, the feedback-induced DW resistance is associated with the DW dynamics. The peculiarity of using a cylindrical nanowire is that the threshold current to initiate the DW dynamics is technically zero~\cite{ref:cylinder}. So, what we mean by DW resistance refers to the difference in $R$ when comparing the results of $I-V$ measurements between a freely moving DW and a pinned DW on identical cylindrical nanowires under the same voltage drop.

The key to understand why such difference is negative for $\beta>2\alpha$ lies in the reaction SMF that propels the electrons along the direction of the DW motion. It contradicts the case of an electric motor where the back emf induction opposes the driving current and raises the system resistance. At the same time, we need to justify that such an anomalous feedback effect does not violate any fundamental physical law. To this end, we make a heuristic analogy between the current-induced DW dynamics in cylindrical nanowires and a water turbine with constant pump, where the rotating wheel represents our moving DW. In fact, the linear velocity of the DW is proportional to its angular velocity, and their ratio is independent of the current~\cite{ref:cylinder}. Therefore, it is equivalent to characterize the DW motion by its angular velocity, which is more transparent to compare with a turbine wheel. Drawing such an analogy is to show that a negative resistance is not surprising, while the analogy itself is by no means exact.

As schematically illustrated in Fig.~\ref{fig:machine}, the working mechanism of a water turbine is compared with an electric motor. They have one thing in common: the steady-state angular velocity $\omega$ increases with decreasing load. So by controlling the load, one can tune $\omega$ in both cases. However, the feedback mechanisms in the two cases are remarkably different. In an electric motor, if one raises $\omega$ by reducing the load, the back emf induced by Faraday's effect will get larger, which counteracts the applied voltage more strongly and reduces the overall current. Consequently, the resistance read off from the $I-V$ curve increases. This realizes the usual negative feedback effect and respects the Lenz law since $I$ decreases when the motor rotates faster. In sharp contrast, if one increases $\omega$ of a water turbine, the water flows more easily in the pipe as the turbine blades less block the water. As a result, the ``resistance" of the entire turbine system appears to be smaller. This feature marks an anomalous feedback: the water current increases when the turbine rotates faster. Ignoring the mass and friction of the wheel, the maximum achievable angular velocity (in the limit of zero load), hence the maximum water current, is set by the water flow in the absence of the turbine. Now go back to our DW dynamics: reducing the DW pinning corresponds to reducing the load on a water turbine, which enhances the driving current in just a similar way as the enhancement of water flow.

Finally, we comment on why the anomalous feedback is more likely to occur in one dimensions. Since $\alpha,\beta\ll1$, the second term of Eq.~\eqref{eq:rkl} dominates the first term, and its coefficient is unlikely to flip sign unless $\beta$ is greater than unity. However, in higher dimensions, the second term always exist, so the first term that could lead to the anomaly is suppressed. Although the second term only refers to the transverse components of the transport, the boundary conditions on the edges can considerably complicate the effective value of the longitudinal component and obscure the observation.

\begin{acknowledgments}
 The authors are grateful to A. Brataas for insightful discussions. We also thank J. Xiao and M. W. Daniels for useful comments. This study was supported by the U.S. Department of Energy, Office of BES, Division of MSE under Grant No.~DE-SC0012509.
\end{acknowledgments}

\appendix*
\section{Derivation of Eq.~\eqref{eq:Linv}}

If all channels are in open circuit conditions except for a particular channel $k$, only the current $J_k$ is nonzero even in the presence of all $N$ thermodynamic forces $X_1\cdots X_N$. The energy dissipation rate is then
\begin{align}
\dot{U}_k=-J_kX_k=-L_{kk}X_k^2-\sum_{i\neq k}^NL_{ki}X_iX_k, \label{eq:dissip}
\end{align}
where the first term is the usual dissipation term. We now eliminate those cross terms $X_iX_k$ ($i\neq k$) in terms of $X_k^2$. Since all currents but $J_k$ are zero, multiplying $X_k$ on $J_i=L_{ij}X_j$ with $i\neq k$ gives:
\begin{widetext}
	\begin{align}
	\begin{bmatrix}
	L_{11} & L_{12} & \cdots & L_{1,k-1} &\ & L_{1,k+1} & \cdots & L_{1N} \\
	\vdots & \vdots & \ddots & \vdots &\ & \vdots & \ddots & \vdots \\
	L_{k-1,1} & L_{k-1,2} & \cdots & L_{k-1,k-1} &\ & L_{k-1,k+1} & \cdots & L_{k-1,N} \\
	&  &  &  &  &  &  & \\
	L_{k+1,1} & L_{k+1,2} & \cdots & L_{k+1,k-1} &\ & L_{k+1,k+1} & \cdots & L_{k+1,N} \\
	\vdots & \vdots & \ddots & \vdots &\ & \vdots & \ddots & \vdots \\
	L_{N1} & L_{N2} & \cdots & L_{N,k-1} &\ & L_{N,k+1} & \cdots & L_{NN}
	\end{bmatrix}
	\begin{bmatrix}
	X_1X_k \\ \vdots \\ X_{k-1}X_k \\ \\ X_{k+1}X_k \\ \vdots \\ X_NX_k
	\end{bmatrix}
	=-X_k^2
	\begin{bmatrix}
	L_{1k} \\ \vdots \\ L_{k-1,k} \\ \\ L_{k+1,k} \\ \vdots \\ L_{Nk}
	\end{bmatrix}.
	\end{align}
\end{widetext}
The coefficient matrix consists of the remaining elements of $L$ after taking away the $k$-th row and the $k$-th column. Regarding the Cramer's rule, the cross term is solved as
$X_iX_k=X_k^2\frac{\mathcal{A}_{ki}}{\mathcal{A}_{kk}}$ for $i\neq k$, where $\mathcal{A}_{ij}$ is the $i,j$-th algebraic cofactor (minor) of $L$. Inserting this relation into Eq.~\eqref{eq:dissip}, and considering the identity of row expansion $\det[L]=\sum_{i=1}^{N}L_{ki}\mathcal{A}_{ki}$, we finally obtain
\begin{align}
\dot{U}_k&=-\left[L_{kk}+\frac{\det[L]-L_{kk}\mathcal{A}_{kk}}{\mathcal{A}_{kk}}\right]X_k^2=-\frac{X_k^2}{[L^{-1}]_{kk}}, \notag
\end{align}
which proves Eq.~\eqref{eq:Linv}.


\begin{thebibliography}{20}
\bibitem{note0} The swapped convention is easy to describe current-induced torques and SMFs. For voltage-induced torques, however, the standard convention is more suitable.
\end{thebibliography}
\end{document}